\documentclass[sigconf]{acmart}
\pdfoutput=1
\usepackage{booktabs} 
\setcopyright{none}
\begin{document}

\acmDOI{}

\acmISBN{}
\acmPrice{}

\acmConference[KDD'17 FATML Workshop]{ACM SIGKDD International Conference on Knowledge Discovery and Data Mining}{August 2017}{Nova Scotia, CAN} 
\acmYear{2017}
\copyrightyear{2017}

\title{The Authority of "Fair" in Machine Learning}
\author{Michael Skirpan}
\affiliation{
  \institution{University of Colorado Boulder}
  \city{Boulder}
  \state{CO}}
\email{michael.skirpan@colorado.edu}
\author{Micha Gorelick}
\affiliation{
  \institution{Fast Forward Labs}
  \city{Brooklyn}
  \state{NY}}
\email{micha@fastforwardlabs.com}

\renewcommand{\shortauthors}{Skirpan and Gorelick}

\begin{abstract}
    In this paper, we argue for the adoption of a normative definition of fairness within the machine learning community. After characterizing this definition, we review the current literature of Fair ML in light of its implications. We end by suggesting ways to incorporate a broader community and generate further debate around how to decide what is fair in ML.
\end{abstract}

\maketitle
\section{Introduction}

The recent boom of machine learning (ML) applications has just as quickly given rise to a slew of critics pointing out the harmful capabilities of these systems. In particular, concerns of bias and discrimination are being debated as ML systems for natural language processing \cite{bolukbasi_man_2016}, judicial sentencing \cite{kirchner_machine_2016}, target advertising \cite{sweeney_discrimination_2013, datta_automated_2015}, image classification \cite{noauthor_google_2015}, and facial recognition \cite{buolamwini_how_2016,timberg_racial_2016} have all proven their ability to inherit bias and create disparate treatment across groups. Responding to these findings is a body of work that attempts to import considerations of "fairness" into our ML approaches \cite{albarghouthi_fairness_2016, celis_how_2016, jabbari_fair_2016, joseph_rawlsian_2016}.

In this paper, we argue for expanding and deepening our approach to "fairness" in ML practice. Drawing from philosophy and ethics, we offer up a normative account of fairness where "fair" is a property that is both communally derived and context dependent. Using this definition, we highlight three categorical framings through which one may inquire about the "fairness" of an ML system: fairness of a system, fairness of an approach, and fairness of a result. We justify these different framings and then move on to overview contemporary approaches to fairness in the ML literature. We argue the position that the literature has thus far focused on the problem of disparate treatment without much attention to other framings. Making this salient allows us to consider the importance of critically examining "whose version of fair" we privilege in ML moving forward and argue that taking a stance on fairness must be understood as invoking an authority that could be more or less legitimate. We conclude by offering possible pathways forward for the community to broaden its approach to fair ML.

\section{The Construction of "Fairness"}

There was once a time when fairness or "the good" was believed to be a static and essential property that could be derived from divine \cite{plato_republic_1945} or rational principle \cite{de_spinoza_ethics_1970}. Determining what was right or fair was a privilege vested in certain authorities and the results were absolute, allowing for no disagreement. This traditional view has changed as modern philosophy and ethics has revised our typological ways of thinking into a normative framing where concepts such as "fair" or "just" are no longer static; rather, they are developed relative to particular communities (who) and contexts (when/how). This is why for modern philosophers such as Richard Rorty \cite{rorty_contingency_1989} or Chantal Mouffe \cite{mouffe_deliberative_1999}, disagreement is part and parcel of a democratic society where we must navigate these tensions in hopes of finding places of commonality from which to move forward.

We are now in an age of {\it machine action} where algorithms can benefit some individuals but may do so at the cost of harming others. Thus, we must not take the responsibility of implementing fair ML systems lightly. Transitioning fragile and contentious matters of human judgment to trained models must be done with care and forethought. As things stand, any engineer with a data set may codify a notion of fairness into an ML system without allowing for any disagreement or community consensus. In order to avoid slipping back into an essentialist morality where a small elite group decides what makes a machine action fair with no recourse, it is critical we expand the available framings and considerations of "fairness" in ML.
 
Taking the stance that we should apply our modern ideas of fairness to ML, we offer the following proposition:

{\bf Proposition (P1):} A machine learning system can only be fair with a contextual justification for the choice of a fairness construct and offering a channel for affected parties to actively assent or dissent to the fairness of the system. 

In P1, a "fairness construct" is a definition for what fair is taken to mean in the problem space and the approach used to codify and measure that definition in training and application. What P1 implies is that a team implementing an ML system should have a sense of the viewpoints around fairness in a domain and be prepared to take a stance when choosing an approach. Further, it dictates some mode of disagreement be available. For example, either a) creating a method of algorithmic due process \cite{crawford_big_2013} where the fairness of a result can be scrutinized or b) working with an active community to address and resolve disputes over time.

A real-world example where we may require such a definition of fairness comes around building classifiers for predicting mental health issues using social media data. In research, there is a tacit acceptance that using social media data to predict mental health is fair \cite{saha_characterizing_2017, manikonda_modeling_2017}. Models and outputs for this have been peer-reviewed and the system itself appears to pass as fair, or at least acceptable. Adopting the normative standard set by this work, one may believe it is fair for anyone to build a similar system to predict mental health, no matter what. And this is exactly what Facebook did--it was classifying teenagers by their psychological vulnerabilities such as feeling "insecure," "worthless," or "needing a confidence boost" \cite{solon_this_2017}. However, Facebook's practice caused a lot of backlash from its users. The reaction to what such a system looks like in practice should raise a red flag that our questions about fairness must go beyond "was the technical approach fair" or "are the results are fair."  It is our belief that the community of ML researchers and practitioners should also be asking questions such as "in what context?", "with what dataset?", and "with what objective?" should a system be trained to classify mental health. It is here we might consider that the discussion of Fair ML cannot be constrained to whether the classifier responds equally to similar inputs. It is our position that ML practitioners must not skirt responsibility in questioning the ethics of what they build so long as some minimum equality criterion is met.

\subsection{Contextualizing Fairness}

In order to clarify our position on the role fairness should play in ML, we offer up three categorical questions one might use to frame an inquiry around the fairness of an arbitrary ML system: {\bf(Q1)} "Is it fair to make X ML system?"; {\bf(Q2)} "I want to make X ML system, is there a fair technical approach?"; and {\bf(Q3)} "I made X ML system, are the results fair?". Each question requires a different set of considerations to arrive at an answer. 

Q1 is asking whether a particular problem, in general, is fit to be approximated or decided upon by an ML system. Using P1, this kind of question requires us to consider the sentiments of the communities it would effect and whether we have a sense of what a fair automation of the task would look like. An immediate response may be, "well anything is worth a try if there's a reasonable data set and approach," but we will come back to why that is a specious assumption. 

Q2 relies on a series of methodological questions perhaps best suited for experts. The fairness of an approach might rely on knowing what a good sample space might look like and the potential biases in historical samples. Answering this question would require an inquiry around whether the set of available features is a close and fair approximation of what we want to predict. Finally, we would hope a practitioner can weigh in on different trade-offs needing consideration for the choice of a training regime (e.g., \cite{zliobaite_relation_2015, kleinberg_inherent_2016}).

Q3 gets at the question of treatment. That is, do the outputs of the algorithm actually correspond to what would constitute a fair response by a human. Answering Q3 requires us to run tests against the model and unpack the algorithm to determine qualities such as what input features were considered important, whether or not different groups have equal chance for misclassification, and whether any variables were acting as proxies for protected variables.

In light of these multiple framings, we take the position that no single framing nor problem space should dominate the realm of what we may consider fair ML practice. While we recognize there will always be organizational ethics that must be further considered beyond what an ML practitioner can influence, we side with an interpretation of engineering ethics similar to that espoused by Langdon Winner \cite{winner_artifacts_1980} that the technology artifact itself has politics and is thus appropriate for normative evaluation. That is, he rejects that all ethics around technological artifacts are socially determined and argues that history has shown us that the artifacts themselves carry political and ethical weight. We now offer a few reasons why the advent of machine learning may embolden such a stance.

\subsection{Why "Fair" Matters}
In line with Winner's position, we should not see an ML model as a blank slate that can only be evaluated after appropriation by some organization. Not only might an ML artifact mediate ethically charged situations, but further it is an artifact carrying some amount of agency. From this vantage, "fair" ML is a recasting of the very idea of fair action in the human sense. Though it may be perceived that adding a "human in the loop" could solve our issue of ethical machine autonomy, research shows that "users may be prone to place an inordinate amount of trust in black box algorithms that are framed as intelligent" \cite{springer_dice_2017}. Meaning that even when an ML model is not acting autonomously, it is causing normative sway that is not neutral.

Further, we must consider that as actions coordinated by an ML model intervene in more of our lives, these actions are not always welcomed or requested \cite{tufekci_algorithmic_2015, andrejevic_big_2014, bilogrevic_if_2016}. As Frank Pasquale points out in "The Black Box Society," how we are categorized through data affects how we will be treated \cite{pasquale_black_2015}. Grounding this thought, we ask whether someone who has never requested therapy, counseling, or any sort of diagnosis should be considered "open game" for a mental health classification. There is not an easy answer to this which begs the original question of whether any ML model that classifies mental health on social media data is fair.

Finally, we want to point to the fact that often the choice of a training objective is contentious in and of itself. There are some cases where the objective has a clear consensus, say in the case of classifying radiology images by whether a cancerous tumor is observed. There is little dispute that the goal here is to be as accurate as possible at identifying cancer. On the other hand, determining whether an individual convicted of a crime might be a repeat offender is likely to solicit a lot of disagreement. Re-appropriating terminology from Richard Rorty \cite{rorty_philosophy_2009}, we distinguish between these two cases as {\it normal} and {\it abnormal} objectives, respectively. Meaning the objective of some ML tasks have a very clear grounds for consensus (normal) and others are highly disputable (abnormal). Due to the fact that ML could be applied to either kind of task, we believe this consideration elevates the level to which an ML model may be considered political or up for normative evaluation.

\section{Current State of Fair ML}

In light of our argument that questions of fairness operate contextually and that the advent of ML elevates our need for normative evaluation of technology artifacts, we move on to apply our contextual framings of fair to the contemporary Fair ML literature. We believe progress has been made, but mostly within the scope of framings Q2 and Q3 and almost exclusively employing some variant of "preventing disparate impact" as the definition of fairness.

\subsection{Is it Fair to Make X ML system?}
Outside of broad critical reflections on the use of technical systems \cite{browne_dark_2015, galloway_laruelle:_2014}, there is yet to be much work characterizing grounds for why an ML system may or may not be a fair approach for a particular problem space. One might anticipate, given a normative definition of fairness, that questions of whether any data could ever approximate certain problems would provide grounds for healthy cross-disciplinary debate. However, "If we can, we ought to," is often treated as an unstated premise for technological development. An exceptional case, is the contemporary debate around autonomous weapons where thousands of scientists have supported a total suppression of development in this area \cite{gibbs_musk_2015}.

In the Fair ML literature, a recent example of researchers asking more general questions about fairness is \cite{bird_exploring_2016}'s evaluation of the ethical implications of ML for autonomous experimentation. Recognizing that practitioners have largely ignored established human-subject research guidelines laid out by the Common Rule and Belmont Principles, the authors argue that automated experimentation may cause harms from privacy-violating inferences and exposing users to less-than-ideal outcomes by being part of the experiment. While falling short of actually questioning whether automated experimentation should be allowed at all, they suggest the adoption of an external review process in light of the fact that it's intractable to obtain consent from each user due to the complexity of the systems being tested. One might construe their high-level question as "Is it fair to make automated experimenting ML systems?" and their answer as "Maybe, but we should have external oversight." 

Though we would be interested in a broader debate about the fairness of research ethics using ML systems, the conclusion drawn supports a more normative evaluation of fairness in this realm. Specifically, the idea that external review may be needed hinges on ideas of authority and context; namely, should we give engineers blanket authority to experiment on users? Akin to P1 above, research ethics are such that a review board should check your experimental standards and participants must be given certain rights. Thus, we agree with the suggestion of external review and further that certain autonomous experiments likely should not be done. Through our lens we would argue many other areas (e.g., ML for biometric inference, ML for emotional persuasion) are ripe for debate around the limitations of what systems are fair to implement.

\subsection{I want to make X ML system, is there a fair technical approach?}

Surveying the ML literature, we see three categorical trends signifying answers this question. The first grouping is research related to interpretability or "white box models" \cite{yang_scalable_2016, ustun_supersparse_2016, wang_falling_2015}. That is, this body of research approaches fairness by developing training methods that aim to produce interpretable ML models. Whether the model is fair becomes a matter of whether an explanation of the results is interpreted as fair. This intersects with the EU's upcoming adoption of "the right to an explanation" law and researchers' calls for algorithmic due diligence \cite{crawford_big_2013}. In our view, this approach has the most affinity with our P1 fairness definition given that it opens up the possibility for models to be interpreted by various subjects (allowing for contextual considerations) and sets forth future possibilities for recourse and disagreement. The limitations of this approach are 1) this approach is not yet feasible for more complex algorithms (ie, most current trends in the ML field) and 2) fairness is bound by the ability for a subject to meaningfully understand and act on the interpretation as the interpretability guarantees nothing about the fairness of the algorithm itself.

Our second categorization includes attempts to resolve disparate treatment concerns by developing statistical independence between predictions and protected categories. These approaches include methods to satisfy fairness by enforcing robust sampling across groups \cite{celis_how_2016}, minimizing the difference in misclassification rates across groups \cite{zafar_fairness_2016}, and training models where protected variables are neither explicitly nor latently used \cite{lum_statistical_2016}. All of these methods define fairness in relation to treatment across known protected classes and give authority to an engineer (or perhaps partnered legal advisor) to {\it a priori} determine which variables are protected. This is a limitation due to the fact that some problem spaces may not have obvious or measurable categories that deserve protection (e.g., should we target someone for prescription drugs at a inferred moment of vulnerability?). A further limitation to this class of approaches shown by ML researchers is that there are inherent trade-offs between a) well-calibrated models, b) parity between groups in the positive class, and c) parity between groups in the negative class \cite{kleinberg_inherent_2016}. 

A third category of work involves decision-making algorithms that attempt to construct fair metrics for optimal choices. These attempts start with the adoption of a quality function that evaluates decisions and then argue for differing approaches toward optimizing choices such as always making choices that minimize regret (ie, integral over time of difference between choice and optimal choice) \cite{joseph_rawlsian_2016} or enforcing that a choice never be made when a higher-quality one was available \cite{jabbari_fairness_2016}. A major limitation of this work is that its baseline standard of fairness is baked into the choice of a quality function, giving a lot of subjective authority to an engineer, and leaving only long-term and short-term optimizations to consider, which can be hard to reason about.

\subsection{I made X ML system, are the results fair?}

This realm of work might be construed as "algorithmic damage control" given that the attempt is not to make a fair algorithm, but rather to develop post-hoc analysis methods that help discover what may be unfair about a black box model. One class of approaches, again premising fairness on prevention of disparate treatment, involves developing mathematical tools to determine whether features related to protected categories (either directly or through co-variates) are influential on the model \cite{adebayo_iterative_2016}. A number of these have come in direct relation to the buzz around the problematic recidivism instrument \cite{chouldechova_fair_2017, yang_measuring_2016}. Given our P1 criterion, these methods, at best, may help an end-user assess whether disparate impact occurred given a set of known protected categories. However, at worst, if we do not know which categories to assess for disparate impact, they may allow a model to pass as fair while unfair biases are still present. In summary, it is a progress that we have ways to verify if an approaches satisfies a disparate impact constraint; however, we are still levied with the challenge of deciding what the constraints should be.

The other major theme in this area of fairness brings back interpretability, but instead of model interpretability, the aim is interpreting why a particular result was obtained \cite{ribeiro_why_2016, ferreira_case_2016}. While this has been successful in certain cases where researchers had an intuition about what kind of internal representation the model may have been using, these interpretations rely on naive, simplified approximations of the model. That is, they are unable to interpret the model globally and instead regress on a set of features in a localized subspace. Again, interpretability satisfies certain normative criteria of fairness by giving some power to an end-user to understand a result. However, so long as we cannot interpret the results globally, it's unclear how much power of recourse one may gain from such an interpretation. Further, this approach only allows us to interpret results using a predefined axis, meaning if we must know what we are looking for before this method becomes useful.

\section{Concluding Remarks}
In the above, we introduced a normative definition of fairness for ML and evaluated it against the current literature. We showed that there are multiple possible framings of fairness that raise different questions about "what is fair in ML" and require different evaluative constructs. Our first summary remark is to point out the limited nature of focusing ML fairness on disparate impact. A corollary of prior research shows that disparate impact and accurate model-making are in fundamental tension \cite{kleinberg_inherent_2016, chouldechova_fair_2017} and ideal fairness defined this way may even be impossible \cite{friedler_impossibility_2016}. Models are discriminators and as such, adding constraints affects one's ability to make a good classifier. Further, we may consider why it is engineers have pressed forward with different disparate impact constructs without yet inviting dialogue with the vulnerable communities for which they are trying to protect.

This brings us to address the most critical takeaway if one accepts a normative criterion of fairness (such as P1): engineers must invite in vulnerable communities and independent advocacy groups to engage in dialogue around fairness. We are sitting on the cusp of a societal transformation where many human intelligent tasks will be transferred to machine intelligence. Exciting as this might be, engineers should be cautious to move too quickly and leave behind the populations who are outside of the networks of the academic and industry elite. If we want to preserve our democratic essence, it is mandatory we develop the standards of the machine through inter-community dialogue. While this conversation is at its beginnings, there are already groups such as the Council for Big Data, Ethics, and Society forming to address these needs. Expanding the number of organizations discussing ethics and working with outside communities, making "usable fairness" a requirement in the development of ML tools, and ensuring that universities and business are teaching fairness and ethics to young engineers will all be critical to legitimizing the authority of the "fairness" embedded in our ML systems.

\bibliographystyle{ACM-Reference-Format}
\bibliography{FATML2017}
\end{document}